\documentclass[aps,prl,twocolumn,preprintnumbers]{revtex4-1}
\usepackage{amsmath, mathrsfs, amssymb,amsfonts,amsthm,graphicx, epsf, dcolumn, yfonts}
\usepackage[hyperfootnotes=true]{hyperref}
\usepackage{color}
\usepackage{slashed}
\usepackage{setspace}
\usepackage{cancel}
\usepackage{wasysym}
\usepackage[lofdepth,lotdepth]{subfig}
\usepackage{float}
\pdfoutput=1
 \newcommand{\be}{\begin{equation}}
 \newcommand{\ee}{\end{equation}}
 \newcommand{\bea}{\begin{eqnarray}}
 \newcommand{\eea}{\end{eqnarray}}

\newcommand{\beq}{\begin{equation}}
\newcommand{\eeq}{\end{equation}}

\renewcommand*{\thefootnote}{\fnsymbol{footnote}}

\begin{document}

\preprint{BRX-TH-6683}

\title{Lorentzian threads as `gatelines' and holographic complexity}
\author{Juan F. Pedraza,$^{\dag,\ddag}$ Andrea Russo,$^{\dag}$ Andrew Svesko$^{\dag}$ and Zachary Weller-Davies$^{\dag}$}
\affiliation{$^\dag$Department of Physics and Astronomy, University College London, London WC1E 6BT, UK\\
$^\ddag$Martin Fisher School of Physics, Brandeis University, Waltham MA 02453, USA}

\begin{abstract}\vspace{-2mm}
 The continuous min flow-max cut principle is used to reformulate the `complexity=volume' conjecture using Lorentzian flows -- divergenceless norm-bounded timelike vector fields whose minimum flux through a boundary subregion is equal to the volume of the homologous maximal bulk Cauchy slice. The nesting property is used to show the rate of complexity is bounded below by ``conditional complexity'', describing a multi-step optimization with intermediate and final target states. Conceptually, discretized Lorentzian flows are interpreted in terms of threads or gatelines such that complexity is equal to the minimum number of gatelines used to prepare a CFT state by an optimal tensor network (TN) discretizing the state. We propose a refined measure of complexity, capturing the role of suboptimal TNs, as an ensemble average. The bulk symplectic potential provides a `canonical' thread configuration characterizing perturbations around arbitrary CFT states. Its consistency requires the bulk to obey linearized Einstein's equations, which are shown to be equivalent to the holographic first law of complexity, thereby advocating a notion of `spacetime complexity'.

\end{abstract}

\renewcommand*{\thefootnote}{\arabic{footnote}}
\setcounter{footnote}{0}

\maketitle

\noindent \textbf{Introduction.} Gravity has an information theoretic character. The sharpest realization of this is captured by the  Ryu-Takayanagi (RT) formula \cite{Ryu:2006bv,Hubeny:2007xt}, relating the area of minimal codimension-2 surfaces $m(A)$ in a $d+1$-dimensional (bulk) AdS spacetime to the entanglement entropy (EE) $S(A)$ of  a conformal field theory (CFT) state restricted to a $d-1$-dimensional boundary subregion $A$ homologous to $m$. The RT formula generalizes Bekenstein-Hawking black hole entropy and satisfies all known properties of the von Neumann entropy. More strikingly, it was used to show gravitational field equations are dual to the first law of entanglement \cite{Lashkari:2013koa,Faulkner:2013ica}, encapsulating the slogan `entanglement=geometry'  \cite{VanRaamsdonk:2010pw,Bianchi:2012ev,Balasubramanian:2014sra}.

Recently, the RT prescription was reformulated in terms of flows or  holographic `bit threads' \cite{Freedman:2016zud}, where $m(A)$ is replaced by the maximum flux of a divergenceless norm-bounded Riemannian vector field $v$ through $A$,
\beq
S(A)=\underset{v\in\mathcal{F}}{\text{max}}\int_{A}\hspace{-2mm}v\,, \;\;\mathcal{F}\equiv\left\{v\,|\,\nabla\cdot v=0\,,\,|v|\leq\tfrac{1}{4G_{N}}\right\}.\label{eq:RTasflowint}
\eeq
The equivalence between the two follows from the max flow-min cut theorem, a known principle in network theory, where the `min cut' is the minimal surface, and was proven using convex optimization techniques \cite{Headrick:2017ucz}. It has several generalizations and applications, \emph{e.g.}, \cite{Harper:2018sdd,Cui:2018dyq,Agon:2018lwq,Harper:2019lff,Agon:2019qgh,Agon:2020mvu,Headrick:toappear,Agon:2021tia,Rolph:2021hgz}. Not only does (\ref{eq:RTasflowint}) have technical advantages, it offers conceptual insight:
 a thread emanating from $A$ is interpreted as a channel carrying a single (qu)bit encoding the microstate of $A$, where the maximum number of threads gives $S(A)$, which may be distilled as Bell pairs.
 Bit threads have also led to insights into TN models of spacetime, \emph{e.g.}, \cite{Vidal:2007hda,Swingle:2009bg}, and the emergence of gravity via the closedness of the `canonical' flow solution
 \cite{Agon:2020mvu}.

Entanglement alone, however, does not describe all aspects of bulk gravitational physics \cite{Susskind:2014moa}. In particular, the late time growth of the Einstein-Rosen bridge inside eternal black holes is not captured by entanglement, but rather complexity. By complexity, one typically means the state complexity, \emph{i.e.}, the smallest number of unitary operators (gates) needed to obtain a particular final state from a given initial state. While the definition of state complexity in a field theory remains an active area of investigation (c.f. \cite{Chapman:2017rqy,Jefferson:2017sdb,Caputa:2017urj}), two proposals for its geometric interpretation have emerged: `complexity=volume' (CV) \cite{Susskind:2014rva,Susskind:2014jwa,Stanford:2014jda,Couch:2016exn} and `complexity=action' (CA) \cite{Brown:2015bva,Brown:2015lvg,Fan:2018wnv}. The CV conjecture says the complexity $\mathcal{C}$ of a CFT state defined on a Cauchy slice $\sigma_{A}$ delimiting a boundary region $A$, so that $\partial A=\sigma_{A}$, is dual to the volume of an extremal codimension-1 bulk hypersurface $\Sigma$ homologous $A$
\beq
\mathcal{C}(\sigma_{A})=\frac{1}{G_{N}\ell}\,\underset{\Sigma\sim A}{\text{max}}\,\text{Vol}(\Sigma(A))\,.\label{eq:CVconj}
\eeq
Here $\ell$ is some undetermined bulk length scale, \emph{e.g.}, the AdS curvature, and
the homology condition $\Sigma\sim A$ implies $\partial \Sigma=\partial A=\sigma_{A}$.
Alternatively, CA equates complexity with the gravitational action $I$ evaluated over the Wheeler-De Witt (WDW) patch.
CV and CA are similar qualitatively, however, here we focus on CV duality.

Given their similar geometric character, it is natural to compare the CV proposal (\ref{eq:CVconj}) to the entropy-area RT prescription.
In light of the bit thread reformulation, one may suspect the CV proposal for holographic complexity (\ref{eq:CVconj}) likewise has a flow based interpretation. Indeed, via the min flow-max cut (MFMC) theorem, where Riemannian flows are replaced by Lorentzian flows, the minimum flux through a boundary region $A$ is equal to the maximum cut of a surface homologous to $A$ \cite{Headrick:2017ucz}. In this letter we use the continuous MFMC principle \cite{Headrick:2017ucz} to reformulate the CV conjecture of holographic complexity in terms of Lorentzian flows and explore some of their properties and implications. We provide a more detailed account and additional results, including explicit geometric realizations of Lorentzian flows in \cite{Pedraza:2021fgp}.


\noindent \textbf{CV and the min flow-max cut theorem.} The continuous version of the MFMC theorem was first presented and proved in \cite{Headrick:2017ucz}. It says the minimum flux of a Lorentzian flow $v$ through a boundary region
$A$ of a compact Lorentzian manifold $\mathcal{M}$ is equal to the volume $V$ of the maximal bulk codimension-1 Cauchy slice $\Sigma\sim A$:
\beq \underset{v}{\text{min}}\int_{A}v=\alpha\,\underset{\Sigma\sim A}{\text{max}}(V(\Sigma)),\;\;\int_{A}v\equiv \int_{A}\sqrt{h}n_{\mu}v^{\mu}\;,\label{eq:mxflmnct1}\eeq
where $\alpha\in\mathbb{R}^{+}$, $n_{\mu}$ is a unit normal covector to $A$, and $\sqrt{h}$ is the induced volume element. The flow $v$ is a timelike vector field obeying,
\beq \nabla\cdot v=0\;,\quad v^{0}>0\;,\quad |v|\geq\alpha\;.\label{eq:Lorflowdef}\eeq

It is now natural to reformulate CV duality (\ref{eq:CVconj}) in terms of Lorentzian flows. Precisely, upon setting  $\alpha=\frac{1}{G_{N}\ell}$, we propose
 $\mathcal{C}$ is the minimum flux of a divergenceless norm-bounded timelike vector field $v$ through $A$
\beq
\!\mathcal{C}(\sigma_{A})=\underset{v\in\mathcal{F}}{\text{min}}\int_{A}\hspace{-1.5mm} v\,, \;\;\mathcal{F}\equiv\left\{v\,|\,\nabla\cdot v=0\,,\,|v|\geq\tfrac{1}{G_{N}\ell}\right\}.\label{eq:cvreformin}
\eeq
Via the MFMC theorem,  (\ref{eq:cvreformin}) is equal to the maximal volume of a Cauchy slice $\Sigma$ homologous to $A$ (\ref{eq:mxflmnct1}).

\noindent \textbf{Properties of Lorentzian flows.}
An important lemma to MFMC  is the nesting property. 
Concretely, consider two nested boundary regions $AB$ and $AB\supset A$, $B\equiv AB\setminus A$, in a compact, oriented Lorentzian manifold $\mathcal{M}$, where $A$ lies to the future of $B$, $A>B$. That is, the boundary is foliated by slices $\sigma_{A}>\sigma_{AB}$. Assuming $\mathcal{M}$ obeys the strong energy condition this foliation induces a foliation of the bulk by non-intersecting maximal cuts $\Sigma(A)>\Sigma(AB)$ \cite{Couch:2018phr}.
Nesting tells us there exists a flow $v(A,AB)$ which simultaneously minimizes flux through $A$ and $AB$. Equivalently, $v(A,AB)$ \emph{maximizes} the flux through $B$, conditioned on minimizing flux through $AB$.

 The nesting property uncovers a number of interesting behaviors holographic complexity must satisfy.
First, when there are two nested regions as above, one has
\beq \mathcal{C}(\sigma_{A})-\mathcal{C}(\sigma_{AB})=-\int_{B}v(A,AB)\;.\label{eq:maxfluxB}\eeq
Since a flow $v(AB)$ with minimal flux through $AB$ has generally less flux than $v(A,AB)$ through $B$, we find
\beq \mathcal{C}(\sigma_{A})-\mathcal{C}(\sigma_{AB})\leq\mathcal{C}(\sigma_{A}|\sigma_{AB})\;,\label{eq:CMI1}\eeq
where we have defined $\mathcal{C}(\sigma_{A}|\sigma_{AB})\equiv-\text{min}\int_{B}v(AB)$.

Since $\mathcal{C}(\sigma_{AB})$ is the complexity of a state at time $t_{AB}=t_{A}-\delta t$, from  (\ref{eq:maxfluxB}) we find the rate $\dot{\mathcal{C}}$ in terms of maximal flux through $B$. By the momentum/volume/complexity (PVC) relation \cite{Susskind:2018tei,Barbon:2020uux,Barbon:2020olv}, we deduce
\beq -\dot{\mathcal{C}}=\lim_{B\to0}\frac{1}{\delta t}\int_{B}v(A,AB)=\int_{\Sigma}T_{\mu\nu}n^{\mu}\zeta^{\nu}-R_{\Sigma}\;.\label{eq:Cdot}\eeq
The first term in (\ref{eq:Cdot}) is the integrated momentum flux $P_{\zeta}$, where $n$ is the future-pointing unit-normal to $\Sigma$ and $\zeta$ is an `infalling' vector  tangent to $\Sigma$ asymptotically equal to a radial, inward-pointing vector with modulus given by the radius of the sphere at infinity. The remainder $R_{\Sigma}$ arises from integrating the momentum constraint and vanishes when $\zeta$ is a conformal Killing vector, in which case the maximal flux through $B$ is only given by  $P_{\zeta}$.

Generalizing to three nested boundary regions $A,AB$ and $ABC$, with $\Sigma(A)>\Sigma(AB)>\Sigma(ABC)$, we uncover the following relationship between the minimal flux $\Phi(X)$ through each region $X$ and complexity $\mathcal{C}(\sigma_{ABC})$
\beq \Phi(AC)+\Phi(BC)-\Phi(C)\leq \mathcal{C}(\sigma_{ABC})\;.\eeq
This is the Lorentzian analog of the strong subadditivity of EE. Moreover, in the limit $B,C$ shrink, together with (\ref{eq:Cdot}), we recover $\ddot{\mathcal{C}}=\dot{P}_{\zeta}$ \cite{Susskind:2019ddc,Barbon:2020olv}, suggesting Newton's laws of gravitation have an origin in complexity.

CV complexity is also known to obey a superadditivity property \cite{Agon:2018zso,Caceres:2018blh,Caceres:2019pgf}, defined in terms of subregion complexity $C_{S}(\sigma_{X})$ \footnote{See the Supplemental Material for details.}, where $\sigma_{X}\subset \sigma_A$ is a boundary spatial subregion. Let $R$ be a Hubeny-Rangamani-Takayanagi (HRT) surface subdividing $\sigma_{A}=\sigma_{X}\cup\sigma_{Y}$. We bipartition $A=A_{X}\cup A_{Y}$ with $A_{X}\cap A_{Y}=\emptyset$. The MFMC theorem allows us to reformulate superadditivity as
\beq C_{S}(\sigma_{X}\cup\sigma_{Y})\geq C_{S}(\sigma_{X})+C_{S}(\sigma_{Y})\;,\eeq
where $C_{S}(\sigma_{X})\leq\int_{A_{X}}\!v(A)$ and similarly for $C_{S}(\sigma_{Y})$.

\noindent \textbf{Interpretation: `gatelines' and tensor networks.}
Similar to the `bit thread' interpretation of Riemannian flows \cite{Freedman:2016zud}, there is a unique mapping between Lorentzian flows and what we call `Lorentzian threads' or gatelines \cite{Headrick:2017ucz}. Specifically, threads are defined as the integral lines of the flows $v$ (\ref{eq:Lorflowdef}) of transverse density $|v|$.
Denoting $N_{A}$ as the number of threads passing through the maximal slice $\Sigma(A)$,
given (\ref{eq:cvreformin}), CV complexity is understood as the \emph{minimum} number of threads passing through $\Sigma(A)$,
\beq \mathcal{C}(\sigma_{A})=\text{min}\; N_{A}\;.\label{eq:minnumthreads}\eeq

\begin{figure}[t]
\includegraphics[scale=0.4]{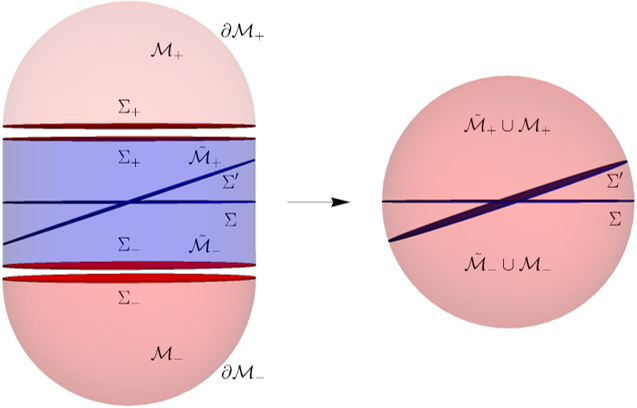}
\centering
\caption{We pick a different bulk Lorentzian slice $\Sigma'$ to partition the two regions, in particular $\Sigma'$ could  be the constant time surface of a Lorentz boosted observer.}
\label{fig: prepforlorentzian2}
\end{figure}

This observation suggests threads prepare the state on $\Sigma(A)$ from a specific reference CFT state defined on the infinite past of the manifold. More precisely, recall bulk Lorentzian spacetimes describe time evolution of CFT  states  prepared by Euclidean path integrals with sources turned on \cite{Skenderis:2008dh,Skenderis:2008dg,Botta-Cantcheff:2015sav,Marolf:2017kvq,Botta-Cantcheff:2019apr}. A reference state is specified on a  bulk Cauchy slice $\Sigma_{-}$ of the southern hemisphere of Euclidean AdS, such that for generic analytic initial data the bulk Einstein's equations reveal which sources are used to prepare the state \cite{Belin:2020zjb}. The length of the Lorentzian cylinder glued at $\Sigma_{-}$ then gives the duration of evolution (Fig. \ref{fig: prepforlorentzian2}).

Threads flowing into $A$ enter from the Euclidean submanifold attached to boundary sources and pass through $\Sigma(A)$. The minimal flux configuration
optimally prepares the CFT state on $\Sigma(A)$, \emph{i.e.}, requiring fewer operations to assemble the state. Thus, Lorentzian threads act as \emph{gatelines}: timelike trajectories representing  unitary gates needed to transform a reference state to a target state. Consequently, complexity is the minimum number of gatelines through $\Sigma(A)$ preparing the target state:
\be\label{Comp-BT}
\mathcal{C}\sim \# \text{ threads} \sim\# \text{ of gates to prepare the state.}
\ee

Conceptually, then, $\mathcal{C}(\sigma_{A}|\sigma_{AB})$ in (\ref{eq:CMI1}) is the minimum number of gatelines needed to prepare a state on $\Sigma(A)$ given the state prepared on $\Sigma(AB)$. That is, $\mathcal{C}(\sigma_{A}|\sigma_{AB})$ is the ``conditional complexity'', describing a two step optimization, first preparing the intermediate state $\Sigma(AB)$ before preparing $\Sigma(A)$. Vanishing flux through $B$ implies the same number of gatelines prepare states on $\Sigma(AB)$ and $\Sigma(A)$, thereby having equal complexity. Meanwhile, when $\mathcal{C}(\sigma_{A})>\mathcal{C}(\sigma_{AB})$, for example, flux through $B$ provides additional gatelines to prepare $\Sigma(A)$.

\begin{figure}[t]
\centering
 \includegraphics[width=3.2cm]{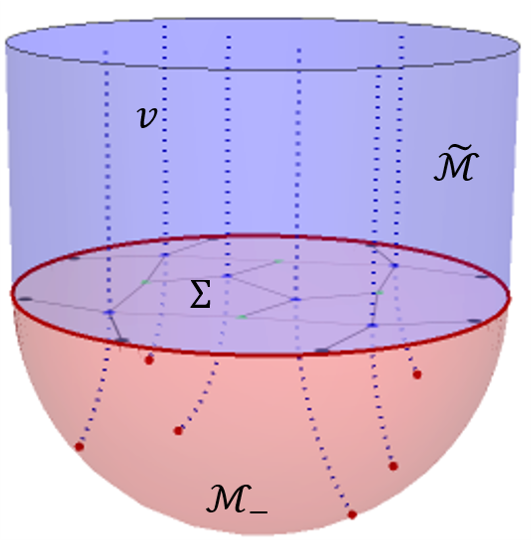}$\qquad\quad$\includegraphics[width=3.2cm]{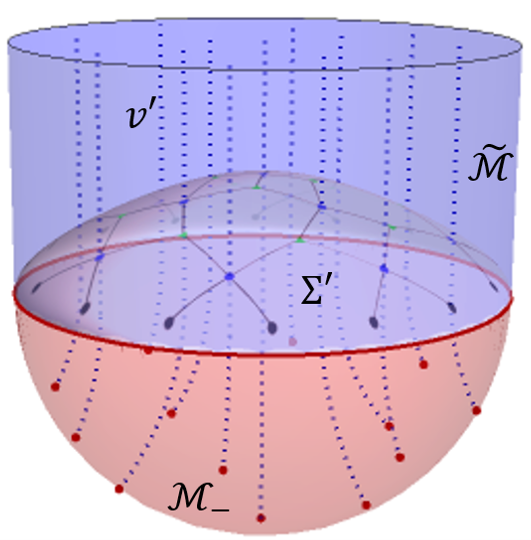}
\caption{Complexity is equal to the minimum number of gatelines preparing a state on maximal volume slice $\Sigma$.
Optimal flow prepares optimal TN (left); suboptimal flows prepare more complex suboptimal TNs (right).
\label{fig:TNs}}
\end{figure}

The gateline interpretation deepens our insight into TN constructions of spacetimes \cite{Vidal:2007hda,Swingle:2009bg}. TNs act as discretizations of bulk spatial slices, where the EE is computed by counting  cuts along the TN, and complexity equals the number of tensors that describes the TN. Combining this  prescription for complexity with (\ref{Comp-BT}), it is then natural to conjecture an optimal thread configuration $v$ prepares the TN on $\Sigma$. We imagine attaching a unitary to each thread, connecting to each physical tensor of the network, so $\# \text{ threads} \sim\# \text{ of tensors}$ (Fig. \ref{fig:TNs}). These unitaries act similar to disentanglers in a MERA TN \cite{Vidal:2008zz}, transforming a reference state to its target. Upon analytic continuation, this operation generates time evolution and the TN acts as a quantum circuit.

\noindent \textbf{Canonical flows from the bulk symplectic form.}
To characterize perturbative excited states, \emph{e.g.}, linear perturbations to vacuum AdS, we develop a notion of `perturbative Lorentzian threads'. Analogous to \cite{Agon:2020mvu},  for a perturbed metric of the form  $g_{\mu\nu}^{\eta}=g_{\mu\nu}+\eta\delta g_{\mu\nu}$, with $\eta$ small, we define $v_{\eta}=v+\eta\delta v+\mathcal{O}(\eta^{2})$, where $v_{\eta}$  obeys the flow criteria (\ref{eq:Lorflowdef}), constraining $\delta v$. Thus, given a metric $g_{\mu\nu}$ and a solution $v$ to the unperturbed min flow problem, we can solve for the minimizing flow $v_{\eta}$.

It is convenient to work with differential forms. We exploit the map between
divergenceless vector fields $v$ and closed $(D-1)$-forms $u$ in a $D$-dimensional background
\begin{equation}\label{eq: flowFormDef}
\begin{split}
&u=\frac{1}{(D-1)!}\epsilon_{\mu_{1}...\mu_{D-1}\nu}v^{\nu}dx^{\mu_{1}}\wedge...\wedge dx^{\mu_{D-1}}\;,
\end{split}
\end{equation}
where $\epsilon = \sqrt{-g} \varepsilon$ is the volume form on the manifold. Criteria (\ref{eq:Lorflowdef}) apply to forms $u$, and we rewrite CV (\ref{eq:cvreformin}) as:
\beq \mathcal{C}(\sigma_{A})=\underset{u}{\text{min}}\int_{A}u\;.\eeq
For metrics perturbatively close to $g_{\mu\nu}$, denote the perturbed $(D-1)$-form by  $u^{\eta} = u + \eta\delta u$. Divergencelessness, the norm bound, and restriction to $\Sigma$ translate to
\beq \label{eq:diffformpertdef}
\begin{split}
&d ( u + \eta\delta u ) = 0 \implies d( \delta u ) =0\;,\\
&- \langle u, u \rangle_g + \eta \left [ 2 \langle u, \delta u \rangle_g + \langle u, u \rangle_{ \delta g} \right]  \geq 1\;,\\
&( u + \eta \delta u ) |_{\Sigma} = \tilde{\epsilon} + \eta \delta \tilde{ \epsilon} \implies \delta u|_{\Sigma} = \delta \tilde{\epsilon}\;,
\end{split}
\end{equation}
with $\langle u,u\rangle_{g}\equiv\frac{1}{(D-1)!}g^{\mu_{1}\nu_{1}}...g^{\mu_{D-1}\nu_{D-1}}u_{\mu_{1}...\mu_{D-1}}u_{\nu_{1}...\nu_{D-1}}$, and $\tilde{\epsilon}$ is the pullback of $\epsilon$ to $\Sigma$. Hence, when studying linear perturbations of complexity around a background we must find a closed $(D-1)$ form $\delta u$ satisfying (\ref{eq:diffformpertdef}).

A canonical choice for $\delta u$ is the bulk Lorentzian symplectic current $\omega^{\text{L}}_{\text{bulk}}$. This follows from the equivalence between the boundary symplectic form $\Omega_{\text{B}}$ and the bulk symplectic form \cite{Belin:2018fxe,Belin:2018bpg}
\begin{equation}
\begin{split}
\Omega_{\text{B}}( \delta_1\tilde{\lambda}, \delta_2 \tilde{\lambda}  )&=
 i \int_{\partial \mathcal{M}_{-} } \omega^{\text{E}}_{\text{bulk}}(\phi, \delta_1 \phi, \delta_2 \phi)\;.
\label{eq: bulktoboundarysymplecticform}
\end{split}
\end{equation}
Precisely, for holographic CFT states prepared by Euclidean path integrals with sources $\lambda$ \cite{Marolf:2017kvq}, the space of sources defines a K{\"a}hler manifold whose K{\"a}hler 2-form $\Omega_{\text{B}}$ is determined by the bulk Euclidean action.
One then invokes the extrapolate dictionary to relate sources $\lambda$ to bulk fields $\phi$, and the variation of the $D$-dimensional Lagrangian form $\delta \mathbf{L}=-E_{\phi}\delta\phi+d\theta(\phi,\delta\phi)$.  Here $E_{\phi}$ is a $D$-form characterizing the equations of motion for $\phi$, which are assumed to be satisfied, $E_{\phi}=0$, and $\theta$ is the symplectic potential, whose variation gives the symplectic current  $\omega_{\text{bulk}}(\phi;\delta_{1}\phi,\delta_{2}\phi)=\delta_{1}\theta(\phi,\delta_{2}\phi)-\delta_{2}\theta(\phi,\delta_{1}\phi)$.

If $\delta_{1,2}\phi$ obey the linearized equations of motion, $\delta_{1,2}E_{\phi}=0$ and $d\omega_{\text{bulk}}=0$, so the southern hemisphere $\partial\mathcal{M}_{-}$ can be pushed to an initial value surface $\Sigma$, replacing $\omega_{\text{bulk}}^{\text{E}}$ with its Lorentzian counterpart $\omega_{\text{bulk}}^{\text{L}}$
\beq \Omega_{\text{B}}(\delta_{1}\tilde{\lambda},\delta_{2}\tilde{\lambda})=\int_{\Sigma}\omega^{\text{L}}_{\text{bulk}}(\phi,\delta_{1}\phi,\delta_{2}\phi)\,.
\label{eq: boundaryEqualsInitial}\eeq
When the boundary sources are deformed by the `new York' transformation $\delta_{Y}$, eq. (3.11) in \cite{York:1972sj}, the bulk symplectic form $\Omega_{\text{bulk}}\!\equiv\!\int_\Sigma\omega_{\text{bulk}}$ is identified with the variation of the volume of the maximal bulk slice $\Sigma$ \cite{Belin:2018fxe,Belin:2018bpg},
\begin{equation}
\Omega_{\text{B}}( \delta_Y\tilde{\lambda}, \delta\tilde{\lambda}) = \Omega_{\text{bulk}}( \delta_Y\phi, \delta\phi) = \frac{(d-1) \tilde{\alpha}}{8 \pi G_{N}}  \delta V\,,
\label{eq:bulksymdeltav}\end{equation}
with $\tilde{\alpha}$ some constant. Foliating the bulk by constant-time surfaces $\Sigma$ with spatial metic $h_{ij}$ and extrinsic curvature $K_{ij}$,
 in terms of $\delta_{Y}$, the  Hamiltonian constraint of general relativity $\mathcal{H}$ reads, $\delta_{Y}\mathcal{H}=2(d-2)K$. This is satisfied when trace of the extrinsic curvature $K=0$. Thence, $\delta_{Y}$ is on-shell when $\Sigma$ is a maximal surface.

For $\tilde{\alpha}\equiv(8\pi/\ell(d-1))$, Eq. (\ref{eq:bulksymdeltav}) naturally proposes a notion of varying complexity, $\delta\mathcal{C}$ \cite{Belin:2018bpg}. In fact, defining complexity as an integral of kinetic energy over the space of sources, $\mathcal{C}(s_{i},s_{f})\equiv\int_{s_{i}}^{s_{f}}ds g_{ab}\dot{\lambda}^{a}\dot{\lambda}^{b}$,
with $s$ parametrizing trajectories in this space, $\delta\mathcal{C}$ obeys a first law
\beq \delta_{\lambda_{f}}\mathcal{C}=\Omega_{\text{B}}(\delta_{Y}\tilde{\lambda},\delta\tilde{\lambda})\;.\label{eq:firstlawC}\eeq
This is a boundary relation. For holographic CFTs, via (\ref{eq: boundaryEqualsInitial}) and (\ref{eq:bulksymdeltav}) one has a first law of (CV) complexity.

Returning to the definition of perturbative thread form $u$ (\ref{eq:diffformpertdef}), it is straightforward to verify $\omega_{\text{bulk}}^{\text{L}}(\delta_{Y},\delta)$ satisfies the conditions on $\delta u$.
Thus $\omega_{\text{bulk}}^{\text{L}}(\delta_{Y},\delta)$ represents a `canonical' thread configuration, solves the MFMC program and is closed for on-shell perturbations.


\noindent \textbf{First law of complexity and Einstein's equations.}
 From (\ref{eq:firstlawC}) we can derive the \emph{covariant} linearized Einstein's equations, differing from \cite{Czech:2017ryf}. Our method is similar to \cite{Lashkari:2013koa,Faulkner:2013ica} using the first law of EE. While the following holds for general bulk states, here we consider vacuum perturbations, hence $\phi$ only represents the bulk metric.

Applying Stokes' theorem and using $E_{\phi}=0$, we have
\begin{equation}\label{eq: integralofdomega}
i \int_{ \mathcal{M}_{-}} d \omega_{\text{bulk}}^{\text{E}}=  \Omega_{\text{B}} ( \delta_Y\tilde{\lambda}, \delta\tilde{\lambda}) - \delta V\;.
\end{equation}
Assuming the holographic version of the first law (\ref{eq:firstlawC}), the right hand side vanishes, requiring $d\omega^{\text{E}}_{\text{bulk}}(\delta_{Y},\delta)=0$ for arbitrary variations $\delta$. Since $\delta_{Y}$ is a diffeomorphism for perturbations around vacuum AdS, then
\begin{equation}
d \omega^{\text{E}}_{\text{bulk}}( \delta_Y, \delta) = - \epsilon \delta E^{\mu \nu} \delta_Y g_{\mu \nu}=0\;,
\label{eq:closedness}\end{equation}
where  $\delta E^{\mu \nu} = \frac{1}{\sqrt{g}}\frac{\delta S_{\text{grav}}}{\delta g_{ \mu \nu}}$,  is no longer a $d$ form on $\mathcal{M}$.

We now argue demanding (\ref{eq:closedness}) for all Lorentzian initial data is equivalent to the linearized Einstein's equations $\delta E^{\mu \nu}=0$ being satisfied everywhere in the bulk $\mathcal{M}$. First consider a \emph{maximal} slice $\Sigma$ along which the southern and northern hemispheres are glued. In Euclidean  Poincar\'e coordinates one has
\begin{equation}
\begin{split}
\hspace{-2.62mm}\tau^2 {\delta E}^{\tau\tau}&+2\tau z {\delta E}^{\tau z}+\left(\tau^2+z^2\right) {\delta E}^{ii}+\tau z^2 {\delta E}^{zz} =0.
\end{split}
\label{eq: numeratorPoincare1}\end{equation}
We now demand this holds for all maximal slices $\Sigma$, each providing data for different Lorentz observers in $\mathcal{M}$, which will allow us to prove $\delta E_{\mu\nu}=0$ everywhere.

We start by deforming the contour to allow for some real time evolution. This is done by gluing a cylinder section $\tilde{\mathcal{M}}$ in between the southern and northern hemispheres, along surfaces $\Sigma_{-}$ and $\Sigma_{+}$ (Fig. \ref{fig: prepforlorentzian2}).
The state on $\Sigma_{-}$ is prepared by a Euclidean path integral over $\mathcal{M}_-$. It then evolves to $\Sigma_+$ and closes at $\mathcal{M}_+$ \cite{Skenderis:2008dh,Skenderis:2008dg}.  Lorentzian AdS $\tilde{\mathcal{M}}$ is split into sections $\tilde{\mathcal{M}}_-, \tilde{\mathcal{M}}_+$ along $\Sigma$. Performing a Wick rotation on $\tilde{\mathcal{M}}_{\pm}$, manifolds $\tilde{\mathcal{M}}_- \cup \mathcal{M}_-$ and $\tilde{\mathcal{M}}_+ \cup \mathcal{M}_+$ describe state preparation on $\Sigma$. However, there is nothing special about $\Sigma$; we could have chosen another slice $\Sigma'$, \emph{e.g.}, a constant-time surface of a Lorentz boosted observer. Wick rotating $\tilde{\mathcal{M}}_{ \pm}^{\prime}$, we have a path integral over the sphere, preparing initial data on $\Sigma'$.

If  $\Sigma'$ is related to $\Sigma$ by an isometry then in the new coordinates the metric will take the same form, as will the initial data on $\Sigma'$. Consequently, $\delta_{Y}$ is invariant, $(\delta_Y g)'(x'(x)) = (\delta_Y g)(x)$. The equations of motion, however, transform as $\delta E^{\prime \mu\nu}(x'(x)) = \Lambda^{\mu}_\gamma \Lambda^\nu_\delta E^{\gamma\delta}(x)$, where $\Lambda^\mu_\nu =\frac{ \partial x^{\prime \mu}}{\partial x^{\nu}}$. Since $d \omega^{\text{E}}_{\text{bulk}}( \delta_Y, \delta)=0$, we deduce
\begin{equation}
\delta E^{\prime \mu \nu} (\delta_Y g)^{\prime}_{ \mu \nu}(x'(x))  = \Lambda^{ \mu}_{ \rho} \Lambda^{ \nu}_{ \sigma} \delta E^{ \rho \sigma} \delta_Y g_{ \mu \nu}(x) =0\;.
\end{equation}
Thus, demanding the constraint (\ref{eq:closedness}) holds for all maximal slices $\Sigma$ in different Lorentz frames means
\begin{equation} \label{eq: moreGeneral}
\Lambda^{ \mu}_{ \rho} \Lambda^{ \nu}_{ \sigma} \delta E^{ \rho \sigma} \delta_Y g_{ \mu \nu} =0
\end{equation}
for any rotation of Euclidean AdS. Together with the Bianchi identities $\nabla_\mu \delta E^{\mu\nu} =0$, this is enough to conclude $\delta E_{\mu\nu}=0$ holds everywhere in $\mathcal{M}$. Hence, assuming CV duality, the first law of complexity implies the linearized Einstein's equations around vacuum AdS.

Note we can easily accommodate higher derivative gravities, with the volume replaced by the `generalized volume' \cite{Bueno:2016gnv}. Also, we emphasize the bulk-boundary symplectic form equivalence (\ref{eq: bulktoboundarysymplecticform}) holds for perturbations over \emph{general} states, not just around vacuum AdS. Then, as suggested in \cite{Lewkowycz:2018sgn} for EE, \emph{any} asymptotically AdS spacetime obeying CV duality and the first law should satisfy the full non-linear Einstein equations. Thus, bulk gravity emerges from boundary complexity.

\noindent \textbf{An ensemble proposal of holographic complexity.}
Thus far we have focused on optimal flows. Here we argue suboptimal flows -- those with more flux and prepare (more complex) suboptimal TNs -- must play a role in defining holographic complexity beyond CV duality.

 We propose a more general prescription using  suboptimal TNs. We are partly motivated by the  `maximin' prescription \cite{Wall:2012uf} for computing EE, a two step algorithm where one picks a slice $\Sigma'$, finds the minimal surface $m_A'$, and then maximizes over all slices.
Extrapolating to TNs,
one first finds the minimal number of cuts on a TN on a $\Sigma'$ and then maximize over all TNs in different $\Sigma'$ \cite{May:2016dgv}:
\be
S_A\sim\underset{\Sigma'}{\text{max}}\,\text{min}\left[\#\text{ cuts}\right]\, .
\label{eq:maximintn}\ee
A lesson is drawn from (\ref{eq:maximintn}): for the computation of EEs, not only the TN on the maximal slice plays a role but also suboptimal TNs defined over other slices. This is particularly crucial for dynamical set-ups.
 Thus, a more refined measure of complexity capturing a notion of state independence is one where all TNs are taken into account. In terms of flows, optimal flows $v$ prepare a TN on the maximal slice while suboptimal flows $v'$ (those with higher flux) prepare TNs over different slices $\Sigma'$. Thus, an averaged measure of complexity accounting all TNs is alternately given by averaging over suboptimal flows.

For specific states, \emph{e.g.}, static ones, the optimal TN is enough to compute the full set of EEs since the associated RT surfaces all lie on a constant-$t$ maximal slice $\Sigma$, and complexity is its volume.  However, for generic out-of-equilibrium settings, $\Sigma$ cannot be foliated by HRT surfaces in general. Consequently, appealing to state independence,
we need to consider TNs defined over all possible slices $\Sigma'$. Since these TNs have different numbers of tensors and thus different complexities, we must consider an appropriate average over $\Sigma'$ to fully characterize the state.
Concretely, we should consider an \emph{ensemble} over all possible TNs defined over all $\Sigma'$. Formally,
\be
\mathcal{Z}\sim\int \mathcal{D}[\Sigma']e^{-\frac{1}{\hbar}\mathcal{S}[\Sigma']}\,,\qquad \Sigma'\in \text{WDW patch},
\label{eq:ensavg}\ee
for a given measure of integration $\mathcal{D}[\Sigma']$ and weight $\mathcal{S}[\Sigma']$ we leave unspecified. We introduced a control parameter ``$\hbar$'' where small $\hbar$ defines a saddle point approximation, where the maximal slice $\Sigma$ emerges as a ``classical'' saddle in the case of static spacetimes, for example, if $\mathcal{S}[\Sigma']\sim\text{Vol}[\Sigma']$. Lastly, $\hbar$ is taken to be a covariant parameter which takes different values depending on the background, \emph{e.g.},
$\hbar$ could be a time-scale of the state, where for static cases $\hbar\to0$, and  $\hbar\neq0$ otherwise.

Assuming (\ref{eq:ensavg}), we propose
\be
\mathcal{C}\sim\frac{1}{\mathcal{Z}}\int \mathcal{D}[\Sigma']\,\text{Vol}[\Sigma']\,e^{-\frac{1}{\hbar}\mathcal{S}[\Sigma']}\,,
\label{eq:genCprop}\ee
for appropriate optimized choices of $\mathcal{S}, \hbar$, and measure of integration. When $\hbar\to0$ we recover CV duality, but generally (\ref{eq:genCprop}) gives a weighted average deviating from CV. Alternatively, in terms of flows we define an average
\beq v_{\text{avg}}\sim\frac{1}{\mathcal{Z}}\int\mathcal{D}[v']v'e^{-\frac{1}{\hbar}\mathcal{S}[v']}\;,\eeq
which obeys $\nabla\cdot v_{\text{avg}}=0$, but relaxes the norm bound.

\noindent \textbf{Discussion.} CV duality reformulated using Lorentzian flows reveals complexity may be interpreted as the minimum number of gatelines needed to prepare an optimal TN discretizing the state, where more complex TNs are prepared by suboptimal flows.
To account for generic TNs
we propose complexity is to be given by a weighted average over all Cauchy slices in the WDW patch.

Our proposal is similar to the holographic dual of the path integral optimization definition of complexity \cite{Caputa:2017urj,Caputa:2017yrh}, where optimization is equivalent to  maximizing an AdS Hartle-Hawking (HH) wavefunction given by a Euclidean path integral of a bulk gravity action over metrics induced on a codimension-1 probe brane $Q$ of tension $T$ \cite{Boruch:2020wax,Boruch:2021hqs}.
Via a saddle-point analysis, the maximization of the HH wavefunction
implies $Q$ provides a constant mean curvature (CMC) slicing of empty AdS.  The tension provides a measure of the complexity: $T\propto K=0$ the path integral complexity functional is optimized; CMC slices $T\neq0$ correspond to suboptimal TNs.
Both proposals thus make use of suboptimal TNs, and we suspect in some contexts the two will coincide. Particularly, when $Q$ foliates the WDW patch the two proposals may be equal when $\Sigma'$ has CMC. Alternatively,  \emph{Lorentzian} path integral complexity was shown to behave as CA duality. It is worth deepening this connection and see how it relates to other complexity proposals \cite{Caputa:2018kdj,Camargo:2019isp,Geng:2019yxo,Chandra:2021kdv}.


\noindent \emph{Acknowledgements.}
We are grateful to C. Ag\'on, J. Barb\'on, E. Caceres, W. Fischler, M. Headrick, M. Heller, T. Jacobson, J. Mart\'\i{}n-Garc\'\i{}a, R. Myers, M. Sasieta, L. Susskind, T. Takayanagi and M. Visser for useful correspondence.  This work is supported by the Simons Foundation via \emph{It from Qubit Collaboration} and by EPSRC.

\bibliography{refs-CBTshort}

\newpage

\appendix

\section*{Supplemental Material}

\noindent \textbf{Notation and conventions.} Here we summarize the various notation and conventions used throughout this letter. Let $\mathcal{M}$ be a $d+1$-dimensional compact Lorentzian manifold in the `mostly plus' signature with boundary $\partial \mathcal{M}$. We are primarily interested in (`bulk') AdS spacetime with a timelike $d$-dimensional conformal boundary, with Euclidean past and future boundaries. Regions of the boundary are denoted by $A,B,..$ \emph{etc.}. Bulk codimension-1 hypersurfaces anchored at the timelike portion of the boundary are denoted by $\Sigma$, and typically represent Cauchy slices. A bulk slice $\Sigma$ homologous to $A$, \emph{i.e.}, $\Sigma\sim A$ obeying $\partial \Sigma=\partial A$, is denoted as $\Sigma(A)$. A necessary and sufficient condition for $A$ to be homologous to purely spacelike $\Sigma$ is that $J^{+}(A)\cap\partial\mathcal{M}= A$, where $J^{+}(A)$ is the causal future of $A$ \cite{Headrick:2017ucz}. Boundary codimension-1 $(d-1)$-dimensional hypersurfaces foliating the timelike portion of $A$ are denoted $\sigma$, with $\sigma_{A}$ reserved as the slice separating $A$ from its complement.

\begin{figure}[t]
\centering
 \includegraphics[width=8.6cm]{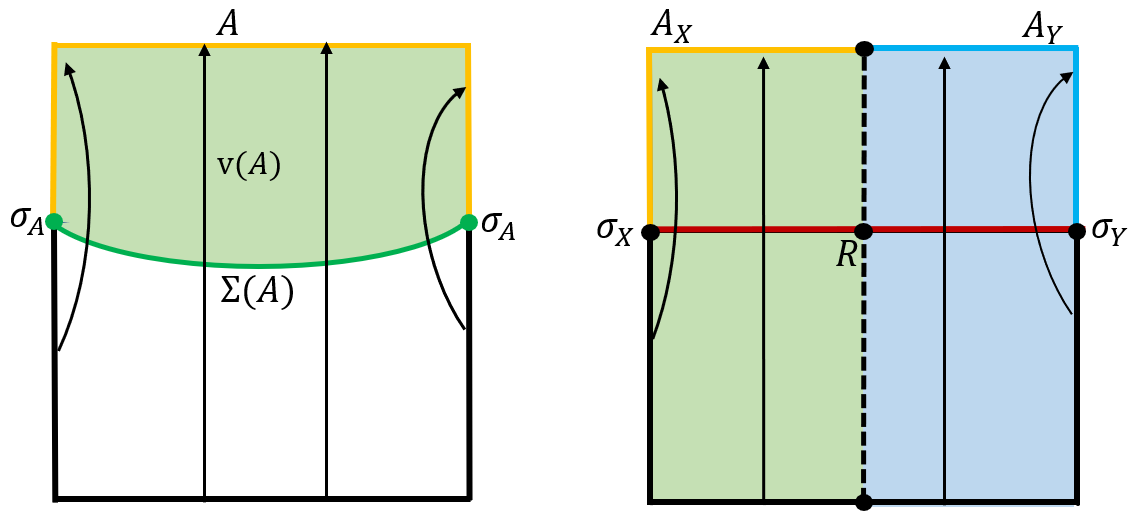}
\caption{Conventions for the computation of holographic complexity of a state on $\sigma_A$ (left) and subregion complexity of the bipartition $\sigma_X\cup\sigma_Y=\sigma_A$ (right).
\label{fig:new}}
\end{figure}

With respect to our notation, the holographic complexity of a CFT state defined on a Cauchy slice $\sigma_{A}$ of a boundary region $A$ is denoted $\mathcal{C}(\sigma_{A})$. Our reformulation of the CV proposal replaces the maximal volume of a hypersurface $\Sigma$ homologous to $A$ with the flux of the minimum flow of a Lorentzian flows $v$ through $A$,
\beq
\!\mathcal{C}(\sigma_{A})=\underset{v\in\mathcal{F}}{\text{min}}\int_{A}\hspace{-1.5mm} v\,, \;\;\mathcal{F}\equiv\left\{v\,|\,\nabla\cdot v=0\,,\,|v|\geq\tfrac{1}{G_{N}\ell}\right\}.\label{eq:cvreforminsupp}
\eeq
We illustrate our conventions in Fig.~\ref{fig:new}. Here $\int_{A}v=\int_{A}\sqrt{h}n_{\mu}v^{\mu}$, where $n_{\mu}$ is the unit normal covector to $A$ and $\sqrt{h}$ the induced volume element. Since $A$ can have both timelike and spacelike sections, $n_{\mu}$ could likewise be timelike or spacelike. However, the divergenceless condition implies $\int_{A}\sqrt{h}n_{\mu}v^{\mu}=\int_{\Sigma}\sqrt{h}n_{\mu}v^{\mu}$, for any $\Sigma\sim A$. For the maximal slice, $\Sigma(A)$, $v^\mu|_{\Sigma}=\alpha n^\mu$ and we recover CV duality $\mathcal{C}(\sigma_A)=\alpha\int_{\Sigma(A)}\sqrt{h}=\alpha \text{Vol}(\Sigma(A))$.



Lorentzian flows $v$ through a boundary region $A$ homologous to any spacelike $\Sigma$ will always have positive flux. This follows from the norm bound and the reverse Cauchy Schwarz inequality for timelike vectors $v^{\mu}$ and $n_{\mu}$, such that $n_{\mu}v^{\mu}\geq|n_{\mu}||v^{\mu}|\geq\alpha$ (evaluated on $\Sigma$).  However, not all Lorentzian flows have positive flux. For example, for two nested boundary regions $A$ and $AB$ ($A\subset AB$), and let $v(A,AB)$ be the simultaneously minimizing flow through $A$ and $AB$. The flux of $v(A,AB)$ through timelike region $B$, however, may be positive or negative. This is a consequence of the fact that, unlike region $A$, the region $B$ obeys $J^{+}(B)\cap\partial\mathcal{M}\neq B$ and hence $B$ is \emph{not} homologous to a spacelike bulk slice $\Sigma(B)$.

\noindent \textbf{Subregion complexity and superadditivity.} The definition of subregion complexity $C_{S}$, previously explored in \cite{Agon:2018zso,Caceres:2018blh,Caceres:2019pgf}, is defined as follows. Let $\sigma_{A}$ be a boundary Cauchy slice of a boundary region $A$ containing spatial subregions $\sigma_{X}$ and $\sigma_{Y}$ with $\sigma_{X}\cup\sigma_{Y}=\sigma_{A}$. Let $R$ be the Hubeny-Rangamani-Takayanagi (HRT) surface dividing $\sigma_{A}$ (where for simplicity we assume the state on $\sigma_{A}$ is pure; if the state is mixed we first purify it such that $\sigma_{X}$ and $\sigma_{Y}$ share the same HRT surface). Then the subregion complexity $C_{S}(\sigma_{X})$ is given by finding the maximal volume slice anchored at $\sigma_{X}\cup R_{X}$ (and similarly for $C_{S}(\sigma_{Y})$. Given the HRT surface for a boundary region, the MFMC theorem provides a flow prescription for $C_{S}$.  In practice, we consider a bipartition of the boundary region $A$ into two non-overlapping boundary regions $A=A_{X}\cup A_{Y}$ (note that here $A_{X}$ and $A_{Y}$ are not nested regions), with associated non-intersecting spatial slices $\sigma_{A_{X}}$ and $\sigma_{A_{Y}}$, where $\sigma_{A_{X}}\cup\sigma_{A_{Y}}=\sigma_{A}$ (see Fig. \ref{fig:new}).  A solution to the min-flow problem for subregion complexity induces a flow $v$ on $\mathcal{M}$ that simultaneously computes the minimal fluxes through spatial slices anchored at $\sigma_{A_{X}}\cup R$ and $\sigma_{A_{Y}}\cup R$. The minimum flux through each slice yields the maximum volumes $V_{A_{X}}$ and $V_{A_{Y}}$, respectively, and is less than or equal to the volume of the maximal slice $\Sigma(A)$. Hence, the subregion complexity of a state reduced to region $\sigma_{X}\cup\sigma_{Y}$, is given by $C_{S}(\sigma_{X}\cup\sigma_{Y})=\int_{\sigma_{A}}v=\int_{A_{X}}v+\int_{A_{Y}}v$. Since the flux of the flow through $A_{X}$ will always be greater than the minimum flux through $A_{X}$, which computes $C_{S}(X)$ (and similarly for $A_{Y}$), superadditivity follows:
\beq C_{S}(\sigma_{X}\cup\sigma_{Y})\geq C_{S}(\sigma_{X})+C_{S}(\sigma_{Y})\;.\eeq


\noindent \textbf{Constructing Lorentzian flows.} Generically, Lorentzian flows are highly non-unique; there are infinitely many timelike vector fields which obey the general criterion of flows. Similar to the Riemannian case \cite{Agon:2018lwq,Agon:2020mvu}, we can develop algorithms to geometrically construct explicit types of Lorentzian flows, as shown in \cite{Pedraza:2021fgp}.

Broadly there are two types of constructions based on (1) integral lines or (2) level sets. In (1), timelike vector fields are found by foliating the bulk Lorentzian spacetime with timelike curves and the norm is fixed using Gauss' law such that the divergenceless condition is satisfied. More precisely, consider a family of integral curves which: (i) are continuous and non self-intersecting;  (ii) are equipped with a tangent vector  $\hat{\tau}$ equal to the unit normal $\hat{n}$ on the maximal volume slice $\Sigma(A)$; (iii) start and end at $\partial\mathcal{M}$. Once a family is picked, we obtain $|v|$ (of flow $v=|v|\hat{\tau}$) such that $\nabla\cdot v=0$ by integrating over an infinitesimal Gaussian pillbox, by requiring
\beq \int\sqrt{h}d^{d-1}x|v|=\text{constant},\label{eq:gauss}\eeq
where $h$ is the induced metric on one of the lids of the pillbox orthogonal to the flow lines. The norm bound is then checked \emph{a posteriori}.
Examples of such flows include ``geodesic flows" built from timelike geodesics foliating the interior of the WdW patch in the given background.

The second algorithm used to explicitly construct flows makes use of a family of level sets which:  (i) contains the bulk maximal slice $\Sigma(A)$ as a member of the family; (ii) are continuous and do not self-intersect; (iii) do not contain closed bulk surfaces, and  (iv) are homologous to $A$. Conditions (i)--(iii) allows one to generate a family of integral lines orthogonal to each hypersurface, from the previous algorithm can be used to construct divergenceless flows. Condition (iv) is not strictly necessary but helps enforce $|v|\geq1$ via the MFMC theorem. Examples of these constructions include so-called ``minimally packed flows" which use maximal volume slices to fix $|v|=1$ everywhere. The timelike Killing flow in AdS spacetimes is an example of such a construction \cite{Couch:2018phr}.


\noindent \textbf{Perturbative flows and the bulk symplectic form.} Motivated by \cite{Agon:2020mvu}, in the main article we developed a notion of `perturbative Lorentzian threads', which are useful to describe linear perturbations to vacuum AdS and correspond to perturbative excited states. 
In this context, it is useful to consider the map between divergenceless vector fields $v$ and closed $(D-1)$ differential forms $u$ in a $D$-dimensional manifold,
\beq v^{\mu}=g^{\mu\nu}(\star u)_{\nu},\;\;(\star u)_{\nu}\equiv\frac{1}{(D-1)!}\sqrt{g}(u\cdot\varepsilon)_{\nu},\label{eq:vucorr}\eeq
where `$\star$' is the Hodge star operator, and $\varepsilon_{\mu_{1}...\mu_{D}}$ represents the Levi-Civita symbol such that $\epsilon=\sqrt{g}\varepsilon$ is the volume form on a manifold with metric $g_{\mu\nu}$. The divergenceless condition $\nabla\cdot v=0$ implies $du=0$ since $du=(\nabla_{\mu}u^{\mu})\epsilon$ via (\ref{eq:vucorr}). Moreover, since $u|_{\Sigma}=(n_{\mu}v^{\mu})\tilde{\epsilon}$, where $\tilde{\epsilon}$ is the pull back of the volume form $\epsilon$ to $\Sigma$, then
\beq\int_{\Sigma}u=\int_{\Sigma}\tilde{\epsilon}=\text{vol}(\Sigma)\;,\eeq
when $\Sigma$ is a maximal volume slice. Thus, by the MFMC theorem, we rewrite the flow reformulation of CV (\ref{eq:cvreforminsupp}) as Eq. (14) in the letter.
We denote the perturbed $(D-1)$-form by  $u^{\eta} = u + \eta\delta u$. The closedness of forms $u^{\eta}$ implies $d(\delta u)=0$, while the maximal volume slice $\Sigma$ is unaltered to leading order in $\eta$ yields $\delta u|_{\Sigma}=\delta\tilde{\epsilon}$.

As we show below, a canonical choice for $\delta u$ is the bulk symplectic current $\omega^{\text{L}}_{\text{bulk}}$. To see this, first note that coherent boundary CFT states $|\lambda\rangle$ are dual to bulk coherent states \cite{Skenderis:2008dh,Skenderis:2008dg,Botta-Cantcheff:2015sav,Marolf:2017kvq}. Specifically, $|\lambda\rangle$ is understood to be a path integral over the boundary of Euclidean $\text{AdS}_{d+1}$, and is the state prepared by initial data on the boundary of the southern hemipshere of the Euclidean manifold, where one inserts sources $\lambda_{\alpha}$ on the boundary. Further note the space of such coherent CFT states defines a K{\"a}hler manifold with a symplectic 2-form $\Omega_{\text{B}}$ \cite{Belin:2018fxe}. Explicitly,
\beq \Omega_{\text{B}}(\delta_{1}\tilde{\lambda},\delta_{2}\tilde{\lambda})=i(\delta_{1}^{\ast}\delta_{2}-\delta^{\ast}_{2}\delta_{1})\log Z_{\text{CFT}}[\tilde{\lambda}]\;,\eeq
where $\tilde{\lambda}$ denote global coordinates on the K{\"a}hler manifold, $\delta_{1,2}$ are variations of the sources, and $Z_{\text{CFT}}[\tilde{\lambda}]\equiv \langle \lambda|\lambda\rangle$ is the partition function of the CFT with sources. For holographic CFTs, the standard AdS/CFT dictionary states $\langle\lambda|\lambda\rangle=e^{-S_{\text{E,grav}}^{\text{on-shell}}[\tilde{\lambda}]}$, where now $\tilde{\lambda}$ set boundary conditions for the bulk fields \cite{Skenderis:2008dh,Skenderis:2008dg}. Thus,
\beq \Omega_{\text{B}}(\delta_{1}\tilde{\lambda},\delta_{2}\tilde{\lambda})=i(\delta^{\ast}_{2}\delta_{1}-\delta_{1}^{\ast}\delta_{2})S_{{\text{E,grav}}}^{\text{on-shell}}[\tilde{\lambda}]\;,\eeq
where $\Omega_{\text{B}}$ denotes the boundary symplectic form for holographic CFTs. Following the steps outlined in the letter, one finds $\Omega_{\text{B}}$ is dual to the bulk symplectic form \cite{Belin:2018fxe,Belin:2018bpg}
\begin{equation}
\begin{split}
\Omega_{\text{B}}( \delta_{1}\tilde{\lambda}, \delta_{2}\tilde{\lambda}  )&=
 i \int_{\partial \mathcal{M}_{-} } \hspace{-2mm}\omega_{\text{bulk}}^\text{E}(\phi, \delta_1 \phi, \delta_2 \phi)\,,
\label{eq: bulktoboundarysymplecticform}
\end{split}
\end{equation}
where $\omega_{\text{bulk}}=\delta_{1}\theta(\phi,\delta_{2}\phi)-\delta_{2}\theta(\phi,\delta_{1}\phi)$ is the bulk symplectic current, $\theta$ the symplectic potential and $\partial\mathcal{M}_{+}$ is the northern hemisphere of the boundary of Euclidean $\text{AdS}_{d+1}$.  When $\delta_{1,2}\phi$ obey the linearized equations of motion, $\delta_{1,2}E_{\phi}=0$, then $d\omega_{\text{bulk}}=0$ and $\partial\mathcal{M}_{-}$ can be pushed to an initial value surface $\Sigma$, replacing $\omega_{\text{bulk}}^{\text{E}}$ with its Lorentzian version $\omega_{\text{bulk}}^{\text{L}}$, Eq. (17) in the letter.

\noindent \textbf{New York deformations and canonical flows.} Consider an ADM decomposition of a Lorentzian manifold $\mathcal{M}$ foliated by constant-$t$ surfaces $\Sigma_{t}$. For Einstein gravity the bulk symplectic form, $\Omega_{\text{bulk}}\equiv\int\!\omega_{\text{bulk}}$, is
\beq \Omega_{\text{bulk}}^{\text{L}}(\delta_{1}\phi,\delta_{2}\phi)=\int_{\Sigma_{t}}(\delta_{1}\pi^{ij}\delta_{2}h_{ij}-\delta_{2}\pi^{ij}\delta_{1}h_{ij})\;.\eeq
Here $h_{ij}$ is the induced metric on $\Sigma_{t}$ and $\pi^{ij}$ is its conjugate momentum. York showed that when one provides initial data on a surface of constant mean curvature (CMC) then the Hamiltonian constraint is solvable by separating the induced metric into a scale $\sqrt{h}$ and a conformal metric $\bar{h}_{ij}=|h|^{-1/d}h_{ij}$ \cite{York:1972sj}. In terms of these new variables, $(\sqrt{h},\bar{h}_{ij})$ and their conjugate momenta, there exists a special deformation known as the `new York' transformation $\delta_{Y}$ \cite{Belin:2018fxe,Belin:2018bpg}
\beq \delta_{Y}h_{ij}=0\;,\quad \delta_{Y}K_{ij}=\tilde{\alpha}h_{ij}\;,\eeq
for some constant $\tilde{\alpha}$, and extrinsic curvature $K_{ij}$.
When the boundary sources on the Euclidean AdS manifold are deformed by $\delta_{Y}$, \cite{Belin:2018fxe,Belin:2018bpg}
showed $\Omega_{\text{bulk}}^{\text{L}}$ is proportional to the change in volume $\delta V$, as in Eq. (18) of the article.

Let us now return to the general description of the perturbative thread form $u_{\eta}=u+\eta\delta u$, for which we recall $d(\delta u)=0$ and $\delta u|_{\Sigma}=\delta\tilde{\epsilon}$. It it is straightforward to verify the bulk symplectic current $\omega_{\text{bulk}}^{\text{L}}(\delta_{Y},\delta)$ satisfies the conditions on $\delta u$, and thus represents a natural (`canonical') thread configuration. To see this, first note that if we perturb initial data using $\delta_{Y}$, then
\beq \omega_{\text{bulk}}^{\text{L}}(\delta_{Y},\delta)|_{\Sigma}=\delta\tilde{\epsilon},\eeq
for maximal slice $\Sigma$. Moreover, since $\delta_{Y}$ is on-shell when $\Sigma$ is maximal, it follows
\beq d\omega_{\text{bulk}}^{\text{L}}(\delta_{Y},\delta)=-\delta E_{\phi}\delta_{Y}\phi=0,\label{eq:OnShellEE}\eeq
assuming on-shell field variations $\delta E_{\phi}=0$. Lastly, we are interested in whether the norm bound $-\langle u^{\eta},u^{\eta}\rangle\geq1$ holds for $\delta u=\omega_{\text{bulk}}^{\text{L}}(\delta_{Y},\delta)$.  We consider instead $\delta v^{\mu}=\delta v^{\mu}_{\eta=0}-\frac{1}{2}v^{\mu}g^{\rho\sigma}\delta g_{\rho\sigma}$, for $\delta v^{\mu}_{\eta=0}g^{\mu\nu}\star(\delta u)_{\nu}$. Motivated by \cite{Agon:2020mvu}, we exploit the fact $\delta v^{\mu}$ depends on the background flow $v^{\mu}$, such that we need only find a single flow $v^{\mu}$ for which $|v_{\eta}|\geq1$. This can be easily checked for variations around vacuum AdS \cite{Pedraza:2021fgp}. Thus, $\omega_{\text{bulk}}^{\text{L}}(\delta_{Y},\delta)$ represents a `canonical' thread configuration that solves the MFMC program and is closed for on-shell perturbations.

\noindent \textbf{Einstein's equations for general backgrounds.} In the letter we explicitly showed that linearized Einstein's equations follow from the first law of (CV) complexity for perturbations around vacuum AdS. It is interesting to ask if this can be generalized to arbitrary backgrounds, in particular, black hole spacetimes. As discussed above, finding the bulk symplectic form is related to the problem of state preparation using the Euclidean path integral. Here one imposes ``initial conditions" on $\partial\mathcal{M}_{-}$ (both normalizable and non-normalizable modes) and then let the bulk equations determine the state on $\Sigma$. One can do this quite generally, including a two sided BH background \cite{Botta-Cantcheff:2019apr}. The new York deformation $\delta_Y$ implements an analogous problem, where one specifies ``boundary conditions" both on $\partial\mathcal{M}_{-}$ and $\Sigma$ and then let the bulk equations of motion determine what sources on $\partial\mathcal{M}_{-}$ are needed \cite{Belin:2020zjb}. Provided the states on $\partial\mathcal{M}_{-}$ and $\Sigma$ are reasonable, one can solve the problem in principle, in which the resulting geometry should be smooth and solves Einstein's equations.  Now, while $\delta_Y$ is not a diffeomorphism in general, for any on-shell background $g$, $\delta_{Y}$ is on-shell, i.e., $\delta E^{\mu\nu} \delta_{Y} g_{\mu\nu} =0$. This will be true whenever the surface $\Sigma$ is maximal.  From Eq. (\ref{eq:OnShellEE}), we then expect the linearized equations of motion should follow quite generally.


\end{document}